\documentclass[twocolumn,showpacs,preprintnumbers,amsmath,amssymb]{revtex4}

\usepackage{tabularx}
\usepackage{array}
\usepackage{mathrsfs}
\usepackage{amssymb}

\usepackage{graphicx}
\usepackage{subfigure}
\usepackage{bm}

\begin{document}

\title{Measurement-device-independent QKD with Modified Coherent State}


\author{Mo Li$^{1,2}$, Chun-Mei Zhang$^{1,2}$, Zhen-Qiang Yin$^{1,2,*}$, Wei Chen$^{1,2,\dagger}$, Shuang Wang$^{1,2}$, Guang-Can Guo$^{1,2}$, and Zheng-Fu Han$^{1,2}$}

\address{
$^1$Key Laboratory of Quantum Information, University of Science and Technology of China, Hefei 230026, China\\
$^2$Synergetic Innovation Center of Quantum Information $\&$ Quantum Physics,University of Science and Technology of China, Hefei, 230026, China\\
$^*$Corresponding author: yinzheqi@mail.ustc.edu.cn\\
$^\dagger$Corresponding author: kooky@mail.ustc.edu.cn
}

\begin{abstract}
The measurement-device-independent quantum key distribution (MDI-QKD) protocol has been proposed for the purpose of removing the detector side channel attacks. Due to the multi-photon events of coherent states sources, real-life implementations of MDI-QKD protocol must employ decoy states to beat the photon-number-splitting attack. Decoy states for MDI-QKD based on the weak coherent states have been studied recently. In this paper, we propose to perform MDI-QKD protocol with modified coherent states (MCS) sources. We simulate the performance of MDI-QKD with the decoy states based on MCS sources. And our simulation indicates that both the secure-key rate and transmission distance can be improved evidently with MCS sources.The physics behind this improvement is that the probability of multi-photon events of the MCS is lower than that of weak coherent states while at the same time the probability of single-photon is higher.
\end{abstract}

\maketitle

\textbf{Introduction.} Quantum key distribution(QKD) empowers two legitimate parties, called Alice and Bob,
to share secret keys even in the presence of a notorious third party, called Eve\cite{BB84}. Theoretically, it's unconditionally secure\cite{security1,security2}. Nevertheless, despite of the proven security in theory, Eve may still be able to crack practical QKD systems by exploiting imperfections in actual implementations\cite{attack1,attack2,attack4,attack7,attack8}. Among these attacks, the detector side channel is the most frequent target. In order to remove all these detector side channel attacks, the measurement device independent QKD(MDI-QKD) has been proposed recently\cite{MDI1,MDI2}, where it is neither Alice nor Bob who detects photons, but some untrusted third party in the middle, Charlie. Alice and Bob can distill secrecy according to Charlie's measurement result of their light pulses. The security of MDI-QKD does not rely on any assumption on Charlie's devices. Recently, a few experimental demonstrations of MDI-QKD have been performed \cite{MDI-EXP1,MDI-EXP3,MDI-EXP4}.
\par Another serious attack arises from the light source. In most experiments, an attenuated laser pulse, called weak coherent state (WCS), is used for substituting the single photon source\cite{MDI-EXP1,MDI-EXP3,MDI-EXP4,qkd1,qkd2,qkd3,qkd4,F-M} which is beyond the present technology. Unfortunately, the WCS source has a Poisson distribution of photon-number and emits n-photon state $|n\rangle$ with the possibility: $P_{\mu}(n)=\frac{\mu^n}{n!}e^{-\mu}$, which means some pulses may contain more than one photon and bring danger to QKD. It is shown that photon-number attack(PNS)\cite{PNS2,PNS3} can allow the eavesdropper to know the secrecy without Alice and Bob's consciousness, largely reducing the secure communication distance. To beat this setback of multi-photon events, decoy state method is
proposed\cite{dtheory1,dtheory2,dtheory3,dtheory4,dtheory5} and has been implemented in many QKD
prototypes\cite{decoyexperiment1,decoyexperiment3,decoyexperiment4,decoyexperiment5,decoyexperiment6,decoyexperiment7}. The principle of the decoy state technique is to estimate the lower bound of incidents caused by single photon which are proven to be absolutely secret. Alice and Bob then distill the corresponding amount of secret information after postprocessing. In decoy state method, the more accurately the bound is made, the higher the key rate can be. And it's apparent that a lower probability of multi-photon events can help improve the performance of decoy state protocols. Actually, there is a kind of modified coherent state source(MCS)\cite{MCS1} which is shown to be capable of eliminating targeted multi-photon number state and can give a lower multi-photon distribution. It has been indicated that this property can benefit the traditional BB84 QKD \cite{MCS2}. In this paper we pay attention to its effect on the MDI-QKD.

\par The MCS source relies on the quantum interference to depress or even eliminate the multi-photon events. The techniques involved are some basic non-linear processes and some linear optical control, all of which are mature nowadays. Its probability distribution can be expressed as\cite{MCS1}:$
|\Psi\rangle_{MCS}=\hat
{\mathcal{U}}|\alpha\rangle=\sum_{n=0}^{\infty}C_n|n\rangle$, where
$\hat {\cal U} =  \exp {1\over 2} (\zeta^*  \hat a^2  - \zeta  \hat
a^{\dagger 2})$,
$C_n = {1\over \sqrt{n! \gamma}} \Big ({\xi\over 2\gamma}\Big)^{n\over
2}\exp\Big({\xi^*\over 2\gamma}\alpha^2 - {|\alpha|^2\over 2}\Big)
H_n\Big({\alpha \over \sqrt{2\gamma\xi}}\Big)$,
$P_n=|C_n|^2$, and
$\gamma \equiv \cosh(|\zeta|), ~\xi \equiv {\zeta\over|\zeta|}
\sinh(|\zeta|),~~ {\rm or}~ \gamma^2= 1+|\xi|^2,\nonumber$
with $\zeta$ being proportional to the amplitude of the pump field. In the above equation, $H_n$ denotes the $n$th-order Hermite polynomial. For MCS source, the mean photon number $\mu$ is calculated by:$\mu=\sum_{n=0}^{\infty}|C_n|^2n$.
\par An interesting character of MCS is that the 2-photon(3-photon) event can be eliminated by setting $\alpha^2=\gamma\xi$ ($\alpha^2=3\gamma\xi$).
Taking 2-photon eliminated(MCS\_2) and 3-photon eliminated MCS(MCS\_3) sources for examples, we list in Table 1 the probabilities of multi-photon events(photon number $n>1$) with the mean photon number $\mu$ being 0.5. The parameters of MCS\_2 are 
$\gamma=1.13252, \xi=0.531601$
, and those of MCS\_3 are 
$\gamma=1.02589, \xi=0.229002$.

\begin{table}
  \centering
  \caption{Multi-photon probabilities and single-photon probabilities of different sources. The mean photon number of the source is 0.5. WCS denotes the weak coherent state source. MCS\_2 denotes the 2-photon eliminated MCS source, and MCS\_3 the 3-photon eliminated. }\begin{tabular}{ccccc} \\ \hline
    Source& single-photon & multi-photon\\ \hline
    WCS &  $0.30326$ & $9.0204\times10^{-2}$\\
    MCS\_2 &  $0.30113$ & $5.7332\times10^{-2}$\\
    MCS\_3 &  $0.37757$ & $5.8606\times10^{-2}$\\
    \hline
  \end{tabular}
\end{table}

It shows that the two MCS sources emit less multi-photon incidents than WCS source. We suppose this lower-multi-photon-probability should decrease the error rate and improve the performance of MDI-QKD protocol, because that multi-photon pulses contribute most of the error counts in MDI-QKD. What's more, in MCS\_3, the single photon's probability is evidently higher that of WCS, which can further benefit the MDI-QKD protocol.
\par\textbf{Simulation Method.}We consider a typical polarization-encoding MDI-QKD as in Ref\cite{MDI1}. However, Alice and Bob are equipped with MCS sources instead of WCS sources. After the quantum communication phase, the data of gain and error rates can be acquired, with which a series of following decoy state equations can be written:

\begin{equation}
Q_{\mu\nu}^{w}=\sum_{n=0,m=0}^{\infty}P_{\mu}(n)P_{\nu}(m)Y_{nm}^w
\end{equation}

\begin{equation}
E_{\mu\nu}^{w}Q_{\mu\nu}^{w}=\sum_{n=0,m=0}^{\infty}P_{\mu}(n)P_{\nu}(m)e_{nm}^wY_{nm}^w,
\end{equation}
where $w$ is the basis choice from either $Z$ basis(rectangular basis) or $X$ basis(diagonal basis). $P_{\mu}(n)$($P_{\nu}(n)$) refers to the probability of n-photon state when the mean photon number is $\mu$($\nu$). $Q_{\mu\nu}^{w}$ and $E_{\mu\nu}^{w}$ are the gain and error rate when both Alice and Bob choose $w$ basis and the mean photon number from them are $\mu$ and $\nu$ respectively. An infinite number of photons need to be dealt with in Eq.(1) and (2), which is an intractable task. In fact, without loss of security and accuracy, certain approximation can be made to make the summation solvable\cite{SIMULATION}. Finally, the key rate of MDI-QKD can be written as\cite{MDI1}:

\begin{equation}
R=P_{\mu}(1)P_{\nu}(1)Y_{11}^z[1-H_2(e_{11}^x)]-Q_{\mu\nu}^zf(E_{\mu\nu}^z)H_2(E_{\mu\nu}^z),
\end{equation}
where $H_2$ represents the binary entropy function, $f(E_{\mu\nu}^z)$ is the reconciliation efficiency, $Y_{11}^z$ is the lower bound of the yield when both Alice and Bob send out single photon of $Z$ basis, and $e_{11}^x$ is the upper bound of the error rate corresponding to incidents that both Alice and Bob send single photon of $x$ basis. With Eq.(1) and (2), we use software-based linear programming method to obtain $Y_{11}^z$ and $e_{11}^x$, such as LinearProgramming function in Mathematica or linprog in Matlab.
\par For simulation, we need to calculate $Q_{\mu\nu}^z$ and $E_{\mu\nu}^z$. Because our goal is to compare the effects of two different light sources on MDI-QKD, the polarization misalignment is not considered. We use the $Z$ basis as an example to explain the principle. Suppose Alice and Bob send their light pulses containing horizontally-polarized $n$ photons $|n\rangle_H$ and vertically-polarized $m$ photons $|m\rangle_V$ respectively, to Charlie's measurement unit. The wave function before Charlie's four detectors, denoted as $D_1^H$,$D_1^V$,$D_2^H$,$D_2^V$, can be derived:
\begin{equation}
\begin{aligned}
&|n\rangle_H|m\rangle_V=\\
&{1\over{\sqrt{2^{n+m}n!m!}}}\sum_{p=0}^n\sum_{q=0}^mi^{n-p+q}{C_n^pC_m^q\over{\sqrt{p!q!(n-p)!(m-q)!}}}\\
&|p\rangle_{D_1^H}|q\rangle_{D_1^V}|n-p\rangle_{D_2^H}|m-q\rangle_{D_2^V},
\end{aligned}
\end{equation}
where $|k\rangle$ represents the k-photon state. For simplicity, both Alice-Charlie's and Bob-Charlie's channel's transmittance efficiency for the photon is assumed to be $\eta$. Then the yields of each subset $|\varphi\rangle_{pq}=|p\rangle_{D_1^H}|q\rangle_{D_1^V}|n-p\rangle_{D_2^H}|m-q\rangle_{D_2^V}$ are calculated as shown in Table 2.

\begin{table*}
  \centering
  \caption{Yields of subset $|\varphi\rangle_{pq}$. In the article the listed gains from top to bottom are abbreviated to $y1$, $y2$, $y3$, $y4$.}\begin{tabular}{ccccc} \\ \hline
    Items & Value \\ \hline
    $y_{D_{11}^{HV}}({pq,nm,HV})$ & $\big(1-(1-\eta)^p+d)\big(1-(1-\eta)^q+d)\big((1-\eta)(1-d))^{n-p+m-q}$\\
    $y_{D_{22}^{HV}}({pq,nm,HV})$ & $\big(1-(1-\eta)^{n-p}+d)\big(1-(1-\eta)^{m-q}+d)\big((1-\eta)(1-d))^{p+q}$\\
    $y_{D_{12}^{HV}}({pq,nm,HV})$ & $\big(1-(1-\eta)^p+d)\big(1-(1-\eta)^{m-q}+d)\big((1-\eta)(1-d))^{n-p+q}$\\
    $y_{D_{21}^{HV}}({pq,nm,HV})$ & $\big(1-(1-\eta)^{n-p}+d)\big(1-(1-\eta)^q+d)\big((1-\eta)(1-d))^{p+m-q}$\\   \hline
  \end{tabular}
\end{table*}

In Table 2 $y_{D_{12}^{HV}}({pq,nm,HV})$ represents the gain caused by the coincidence of $D_1^H$ and $D_2^V$. $pq$ refers to the $|\varphi\rangle_{pq}$ subset and $(nm,HV)$ means the polarizations of $n$-photon pulse, and $m$-photon pulse are horizontal and vertical respectively. $d$ represents the dark count rate in per pulse. From Eq.(4) it can be known that the probability related to this subset is $p_{nm}^{HV}(|\varphi\rangle_{pq})=|{1\over{\sqrt{2^{n+m}n!m!}}}i^{n-p+q}{C_n^pC_m^q\over{\sqrt{p!q!(n-p)!(m-q)!}}}|^2$. Hence the yield of $|n\rangle_H|m\rangle_V$ is, theoretically:
\begin{equation}
\begin{aligned}
&Q_{nm}^{HV}=\sum_{p=0}^n\sum_{q=0}^mp_{nm}^{HV}(|\varphi\rangle_{pq})\Big(y1+y2+y3+y4\Big),
\end{aligned}
\end{equation}
which can further lead to the overall gain when Alice and Bob's mean photon numbers are $\mu$ and $\nu$ respectively:
$Q_{\mu\nu}^{HV}=\sum_{n=0,m=0}^{\infty}P_{\mu}(n)P_{\nu}(m)Q_{nm}^{HV}$, where an infinite summation is encountered. Practically we stop the summation where the photon number from either Alice or Bob is more than 12, which is accurate enough and results in a relative deviation smaller than $10^{-8}$.
\par According to the MDI-QKD protocol, all items listed in Table 2 are legal incidents and contribute to final keys, indicating that no error will be caused in this situation. For the other polarization combinations of rectangular basis: $|V\rangle|H\rangle$, $|H\rangle|H\rangle$, $|V\rangle|V\rangle$, we can follow the same procedure as above to obtain its gain, but note that in $|H\rangle|H\rangle$ and $|V\rangle|V\rangle$ situations, all gains are not legal and responsible for errors. Eventually, the gain and error rate of $Z$ basis with Alice's and Bob's mean photon number being $\mu$ and $\nu$ are (assume that polarizations are chosen randomly):
\begin{equation}
\begin{aligned}
Q_{\mu\nu}^{Z}={1\over4}(2Q_{\mu\nu}^{HV}+Q_{\mu\nu}^{HH}+Q_{\mu\nu}^{VV})
\end{aligned}
\end{equation}
\begin{equation}
\begin{aligned}
E_{\mu\nu}^{Z}={{Q_{\mu\nu}^{HH}+Q_{\mu\nu}^{VV}}\over{4Q_{\mu\nu}^{Z}}}.
\end{aligned}
\end{equation}

, where $1/4$ means the probability of the combination of Alice's and Bob's basis:$\{HV, VH, HH, VV\}$. The treatment for $X$ basis is similar to the procedures above.

\textbf{Simulation.}With the method introduced in the last section, the simulation can be conducted for both WCS source and MCS source. Both MCS\_2 and MCS\_3 are studied here. The results are shown in Fig.1 and Fig.2. The dark counts of single photon detectors are all taken to be $6 \times 10^{-6}$ per pulse and loss coefficient of channel is 0.2dB/km. Let the detection efficiency be $100\%$. The combination of vacuum state$+$decoy state$+$signal state is used, whose mean photon numbers are taken to be 0, 0.1 and 0.5 respectively. The reconciliation efficiency is fixed at $1.2$, which is conservative enough for practical system.



\begin{figure}[htb]
\centering
\subfigure{
\begin{minipage}{\linewidth}
\includegraphics[width=\linewidth]{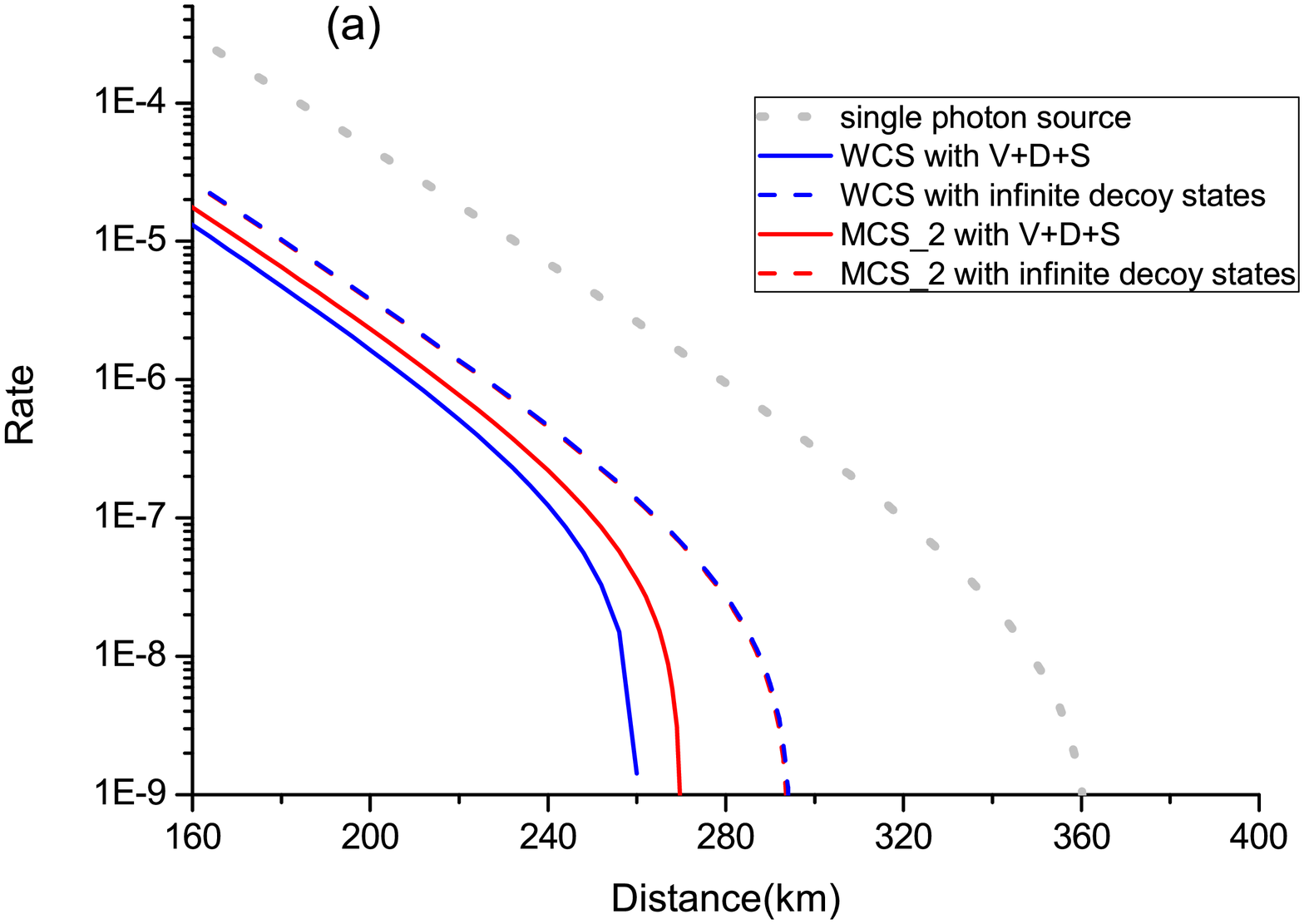}
\end{minipage}
}
\subfigure{
\begin{minipage}{\linewidth}
\includegraphics[width=\linewidth]{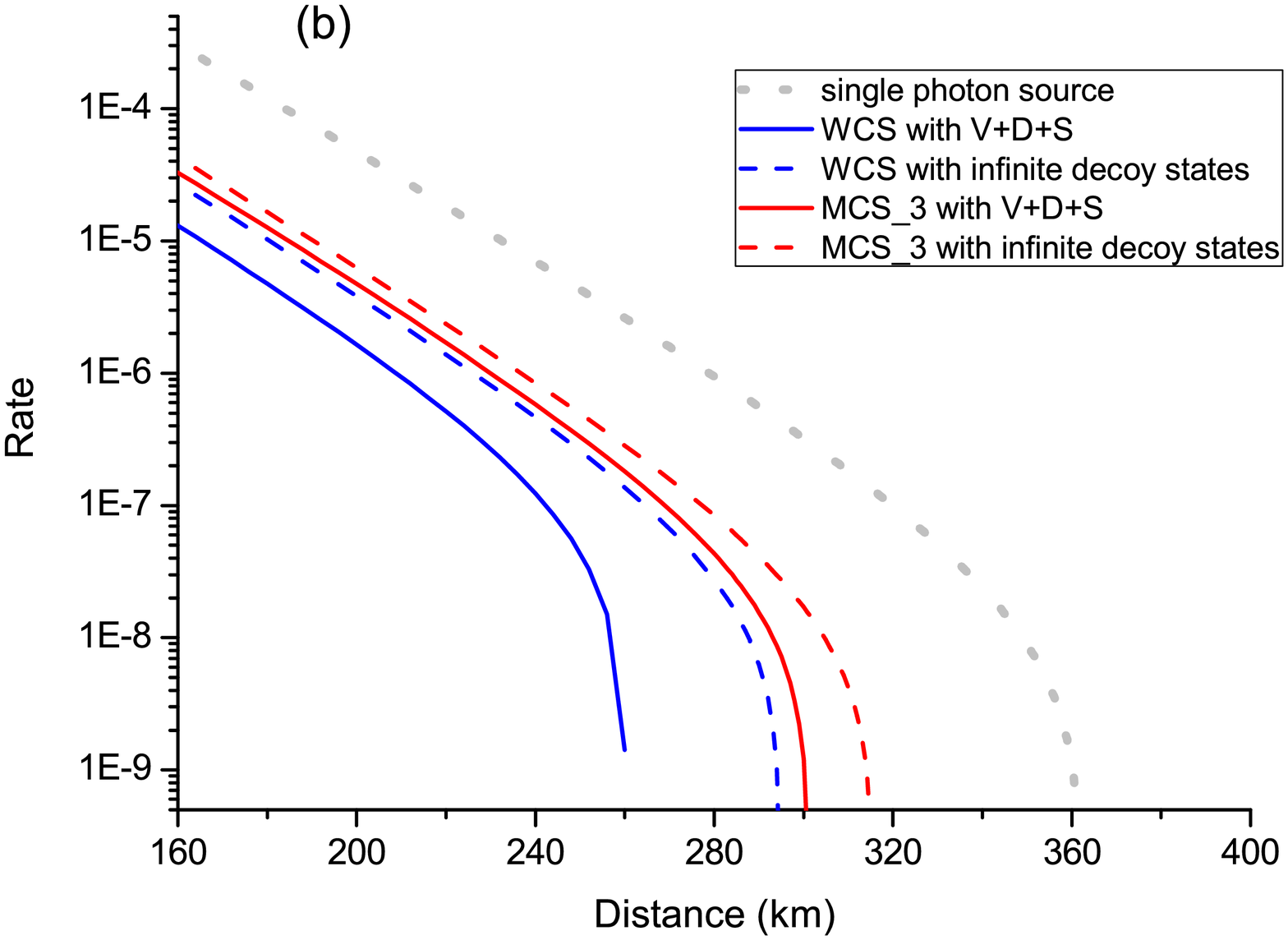}
\end{minipage}
}
\caption{Comparison between WCS source and MCS source. \textbf{(a)}:WCS and MCS\_2. \textbf{(b)}: WCS and MCS\_3. Note that in \textbf{(a)} the lines of infinite decoy states of WCS and MCS\_2 overlap each other.}
\end{figure}

From Fig.1(a) and (b), under the configuration of vacuum$+$decoy$+$signal the communication distance is extended by around 10km and 40km respectively with MCS\_2 and MCS\_3, and the key rate is evidently higher than WCS. As supposed, the lower multi-photon probability can truly raise the key generation-rate. MCS\_3's high single-photon probability also increases the distance evidently.

In the above the simulation is conducted without fluctuation in data, corresponding to the situation of infinite pulses. Now we give a brief research into the finite scenario which would bring fluctuation into the result(Fig.2). Here five standard deviations of fluctuation are used. We assume that the pulse number of vacuum state, decoy state and signal state are the same, denoted by N. As illustrated in Fig.2, the performance varies heavily with different pulse numbers. However, in the linear regime before the cutting-off, the protocol can work at the level close to the no fluctuation situation.
\begin{figure}[htb]
\centerline{\includegraphics[width=7.5cm]{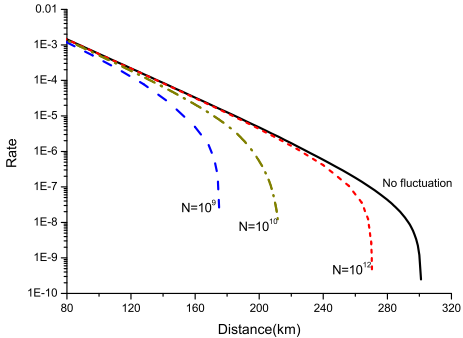}}
  \caption{Key rate with fluctuation introduced using MCS\_3. Five standard deviation of fluctuation is used. N denotes the number of pulses.}\label{infinite decoy}
\end{figure}

 \par In the practical implementation the mean photon number of the decoy state may be different from that used in the simulation above and the number of decoy states may be more than one, for achieving as close as possible to the performance of infinite decoy situation(for example, if the mean photon number of the decoy state is taken to be 0.02, the distance of WCS can stretch to 147km which is almost the same with that of infinite decoy situation). So here we give an investigation of the situation that both WCS and MCS are optimally configured with decoy states, while the signal state's mean photon number is fixed at 0.5. The result is shown in Fig.3.
\begin{figure}[htb]
\centerline{\includegraphics[width=7.5cm]{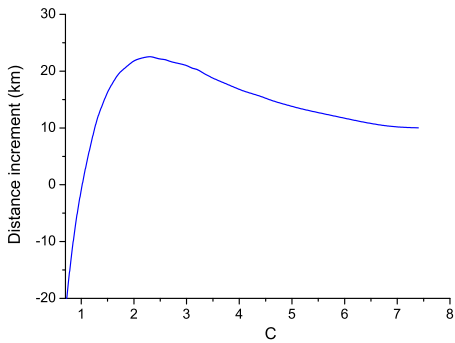}}
  \caption{Distance increment by MCS when MCS and WCS are both best configured with decoy states, which is also the result of implementing infinite decoy states.C is defined as $\alpha^2=C\gamma\xi$}\label{infinite decoy}
\end{figure}

\textbf{Conclusion.}Because that the MCS source can give out a lower probability of multi-photon and higher probability of single-photon, we suppose it should benefit the decoy state technique. The simulation results support our speculation, showing that the MCS source can outperform the WCS source by several dozen kilometers. We think an experiment based on this will be very meaningful.
\par This work was supported by the National Basic Research Program of China (Grants No. 2011CBA00200 and No. 2011CB921200), National Natural Science Foundation of China (Grants No. 61101137, No. 61201239, and No.61205118).
\par We thank Fang-Wen Sun for helpful discussion.

\end{document}